\documentclass[prb,showpacs,twocolumn,superscriptaddress]{revtex4-1}
\usepackage{graphicx,color}
\begin{document}

\title{Exotic phase separation in one-dimensional hard-core boson system with two- and three-body interactions}
\author{Chen Cheng}
\affiliation{Center of Interdisciplinary Studies $\&$ Key Laboratory for Magnetism and Magnetic Materials of the Ministry of Education, Lanzhou
University, Lanzhou 730000, China}
\author{Bin-Bin Mao}
\affiliation{Center of Interdisciplinary Studies $\&$ Key Laboratory for Magnetism and Magnetic Materials of the Ministry of Education, Lanzhou
University, Lanzhou 730000, China}
\author{Fu-Zhou Chen}
\affiliation{Center of Interdisciplinary Studies $\&$ Key Laboratory for Magnetism and Magnetic Materials of the Ministry of Education, Lanzhou
University, Lanzhou 730000, China}
\author{Hong-Gang Luo}
\affiliation{Center of Interdisciplinary Studies $\&$ Key Laboratory for Magnetism and Magnetic Materials of the Ministry of Education, Lanzhou
University, Lanzhou 730000, China}
\affiliation{Beijing Computational Science Research Center, Beijing 100084, China}

\date{July 24, 2014}
\pacs{03.75.Lm, 67.85.Hj}

\begin{abstract}
We investigate the ground state phase diagram of hard-core boson system with repulsive two-body and attractive three-body interactions in one-dimensional optic lattice. When these two interactions are comparable and increasing the hopping rate, physically intuitive analysis indicates that there exists an exotic phase separation regime between the solid phase with charge density wave order and superfluid phase. We identify these phases and phase transitions by numerically analyzing the density distribution, structure factor of density-density correlation function, three-body correlation function and von Neumann entropy estimator obtained by density matrix renormalization group method. These exotic phases and phase transitions are expected to be observed in the ultra-cold polar molecule experiments by properly tuning interaction parameters, which is constructive to understand the physics of ubiquitous insulating-superconducting phase transitions in condensed matter systems.
\end{abstract}

\maketitle

\section{Introduction}

Identifying a variety of phases and the phase transitions between them in a many-body correlated system is one of central challenges in condensed matter physics, for example, the complex phase diagram of hole-doped cuprates,\cite{Damascelli2003, Lee2006} in which an intriguing phenomenon is how the insulating phase of the parent compounds does evolve into superconducting phase with lightly doping. Actually, this is a controversy issue and concerns the essential physics of the pseudogap phenomena in hole-doped cuprates, as intensively discussed in most recent. \cite{Morr2014, Comin2014, Neto2014}

Over the last two decades, with the great advance of the cold atom and/or ultra-cold molecule experiments, besides the interest in its own right in low-temperature physics, these systems have been proved to be able to simulate many model Hamiltonians in condensed matter physics by engineering microscopically optical lattice,\cite{Bloch2008} which provides a significant way to understand the physics of phase transitions. Importantly, unlike the systems in solid state physics, the merit of these cold atom and/or molecule systems is that the interaction parameters can be tuned by applying external controllable fields, which provides an ideal platform to investigate in detail the process of the phase transitons, such as the Mott-Hubbard transition in optical lattices, \cite{Jaksch1998, Kuhner1998} as observed in cold atom experiment. \cite{Greiner2002} This phase transition can be described by a conceptually simple Bose-Hubbard model, \cite{Fisher1989} only containing the kinetic energy and on-site repulsive interaction. The zero temperature phase diagram of this model shows the transition from a series of Mott insulating with fixed integer filling to superfluid phase. \cite{Bloch2008} When the two-body interaction is absent, it has been shown that the Bose-Hubbard model with three-body repulsive interaction also shows a similar Mott insulating to superfluid phase transition. \cite{Sansone2009, Zhou2010, Valencia2011, Valencia2012, Safavi-Naini2012} When both two- and three-body interactions exist and are tuned independently, more novel phases can occur, for example, a dimer superfluid phase have also been reported in systems with two-body attractive and three-body repulsive interactions. \cite{Chen2011, Daley2009, Daley2014}

The experimental realization of the ultra-cold polar molecules further provides a tool to explore the many-body physics with long-range interaction, as proposed by Buchler \textit{et al.} \cite{Buchler2007} In such a system, the polar molecules can be driven by microwave fields in optical lattice to obtain effective hard-core Bose-Hubbard model with dipolar-dipolar three-body interaction, which can be approximated by off-site three-body interaction. Such a three-body interaction also can be obtained by spin-1 system in optical lattice. \cite{Mazza2010, Mahmud2013} The phase diagram of this model in one-dimensional case has been studied by Capogrosso-Sansone \textit{et al.} \cite{Sansone2009} They found that at unconventional filling $n=2/3$ there is a phase transition between solid and superfluid phases.

\begin{figure}[tbp]
\begin{center}
\includegraphics[width=\columnwidth]{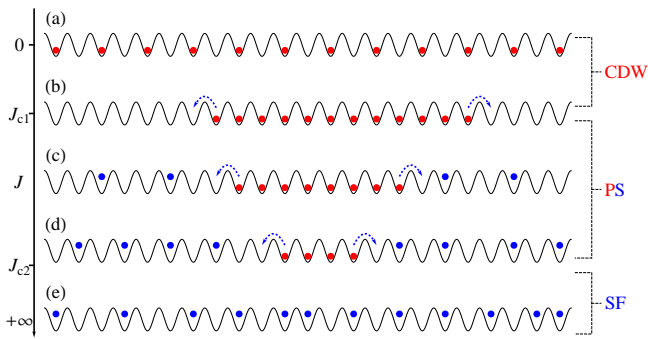}
\caption{(Color online) Schematic phase diagram and the corresponding patterns (a-e) of hard-core bosons with two-body repulsive $U$ and three-body attractive $W$ interactions in one-dimensional optical lattice as a function of the hopping rate $J$. When $W \sim -U$ and increasing $J$, the system experiences the solid phase with charge density wave (CDW) order, the phase separation (PS) and the superfluid (SF) phases. The phase transitions happen around $J_{c1}$ from the CDW to PS phases and around $J_{c2}$ from the PS to SF phases. The red (blue) dots denote the localized (delocalized) bosons. The blue dashed arrows denote the hopping of bosons at two ends of the island.} \label{fig1}
\end{center}
\end{figure}

Motivated by these rich phases and phase transition behaviors, here we explore the interplay of two- and three-body interactions of the hard-core boson system in one-dimensional optic lattice. In particular, we focus on the case of two-body repulsive ($U$)and three-body attractive ($W$) interactions. Such a system can show some exotic phases, as presented schematically in Fig.\ref{fig1} in the case of $U \sim |W|$ for half-filling. When the hopping rate $J$ is zero, intuitively, the bosons fill into the one-dimensional optical lattice like that in Fig. \ref{fig1}(a) denoting the solid phase with charge density wave (CDW) order. In this case, the system has a ground state energy of zero. When the hopping rate is switched on, the bosons trend to hop between sites. However, the two-body repulsion interaction prevents the movement of the bosons but the three-body attractive interaction benefits the clustering of the bosons. If one neglects the kinetic energy, the boson fragments in the optical lattice are obviously unfavorable since each fragment contributes an energy of $U$ neglecting its size. In this case, all bosons like to gather at certain place to form an island with energy $U$, as shown in Fig. \ref{fig1}(b). However, this boson island is also unstable and the bosons at the two ends of it can hop outward to show liquid behavior, which lowers the energy. This is the beginning of the exotic phase separation (PS) phase. Further increasing $J$, the bosons at the two ends of the island continue to hop outward, the size of the island shrinks, as shown in Fig. \ref{fig1}(c) and (d). Finally, when the hopping rate dominates the interactions, all bosons like hopping and the system results in the superfluid (SF) phase shown as Fig. \ref{fig1}(e). In the process, the system experiences two phase transitions, the first is from the CDW to PS phases and the second one is from the PS to SF phases. The existence of the exotic PS phase located between the CDW and SF phases is central result of the present work. In the following we use density matrix renormalization group (DMRG) \cite{White1992, White1993} method to numerically confirm the above picture.

\section{Model Hamiltonian and Numerical Method}

\begin{figure}[tbp]
\begin{center}
\includegraphics[width=\columnwidth]{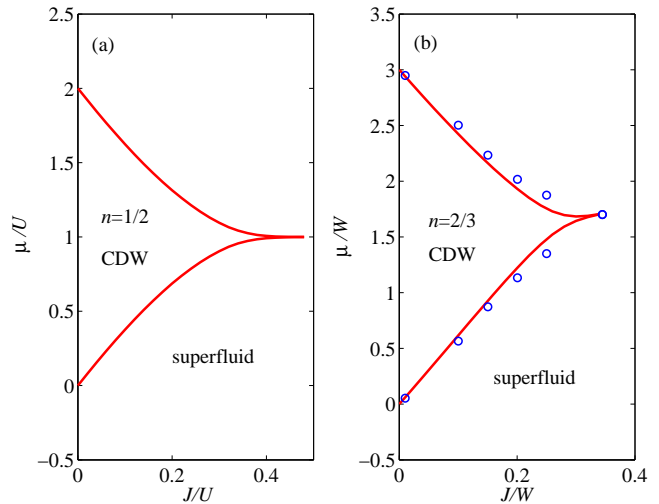}
\caption{(Color online) Ground state phase-diagram of hard-core bosons in one-dimensional optical lattice with (a) two-body repulsion $U$ and (b) three-body repulsion $W$ obtained by DMRG with open boundary condition. The results in the thermodynamic limit have been obtained by finite size scaling analysis with $L = 59, 119$, and $179$. In both cases a phase transition from CDW to SF phases happens. For the two-body repulsion case, the phase transition happens around $J_c/U = 0.48$, which is roughly consistent with the result of $J_c/U = 0.5$.\cite{Cazalilla2011} For the three-body repulsive case, the blue circles are the result of Ref.[\onlinecite{Sansone2009}] by quantum Monte Carlo calculation under temperature $T=0.6J/L$ with periodic boundary condition. Both results indicate that the phase transition happens around $J/W = 0.34$.}\label{fig2}
\end{center}
\end{figure}

The hard-core bosons with two-body repulsive and three-body attractive interactions in one-dimensional optical lattice can be described by the following Hamiltonian
\begin{equation}
H=-J\sum_{\langle i,j \rangle}(b_i^{\dag} b_j + h.c.)+ U\sum_i {n_i n_{i+1}} + W\sum_i {n_i n_{i+1} n_{i+2} },\label{eq1}
\end{equation}
where $b^{\dagger}(b)$ is the bosonic creation (annihilation) operators satisfying the hard-core constraint, and $n_i = b_i^{\dagger} b_i$ is the density operator at site $i$. The first term describes kinetic energy with hopping rate $J$, the latter two terms denote off-site two- and three-body interactions with strengths $U$ and $W$, respectively. Except for special address, hereafter the two-body interaction is repulsive and the three-body one is attractive. Special case of Eq. (\ref{eq1}) with only two-body repulsive interactions ($W=0$) is known to have a ground state phase diagram of solid phase with CDW order and superfluid at half filling.\cite{Cazalilla2011} Recently, a similar phase transition has been also obtained by using quantum Monte Carlo method at unconventional filling $n=2/3$ when the two-body repulsive interaction is replaced by the three-body one. \cite{Sansone2009}
As usual, these phases can be determined by excitation gap in such a finite system
\begin{equation}
\Delta_\mu(L,N) = \mu^p(L,N) - \mu^h(L,N)
\end{equation}
where $L$ is the lattice size and $N$ is the number of bosons, $\mu^p$ ($\mu^h$) is the particle (hole)-like chemical potential, determined as the energy it takes to add a particle (hole) to the system, respectively,
\begin{eqnarray}
&& \mu^p(L) = E_0(L,N+1) - E_0(L,N), \\
&& \mu^h(L) = E_0(L,N) - E_0(L,N-1),
\end{eqnarray}
where $E_0(L,N)$ is the ground state energy for system with $L$ sites and $N$ bosons. The existence of non-zero excitation gap is one of the key character of solid phase.

In the following we numerically solve the Hamiltonian (\ref{eq1}) by using DMRG, which is powerful in dealing with the one-dimensional systems. Here we choose open boundary conditions (OBCs) since it is well known that DMRG is more accurate for open chains than that for periodic boundary conditions. In this case, the boson number is fixed as $(L+1)/2$, \cite{note} which approaches the half filling for large lattice size. The physics can be obtained by finite size scaling, as shown below.

To confirm the validity of our program, we first calculate the cases of only repulsive interactions and compare the results with those reported in the literatures. The results are shown in Fig. \ref{fig2}. The system shows two phases, namely, solid phase with CDW order and superfluid phase. For the two-body repulsion case, the phase transition of the system at half-filling happens around $J_c/U = 0.48$, which is roughly consistent with the result of $J_c/U = 0.5$.\cite{Cazalilla2011} In the three-body repulsive case, the system at filling $n = 2/3$ the phase transition happens around $J_c/W = 0.34$, which is consistent with the result $W_c/J=2.80\pm0.15$ that obtained by quantum Monte Carlo simulation, \cite{Sansone2009} as shown as the blue circles. Noted that in DMRG calculation we use filling $n=2N/(3N-1)$ which is equivalent to $2/3$ in the thermal dynamic limit. The comparison confirms the validity of our DMRG program and finite size scaling.

\section{Exotic phase separation}

\begin{figure}[tbp]
\begin{center}
\includegraphics[width=0.9\columnwidth]{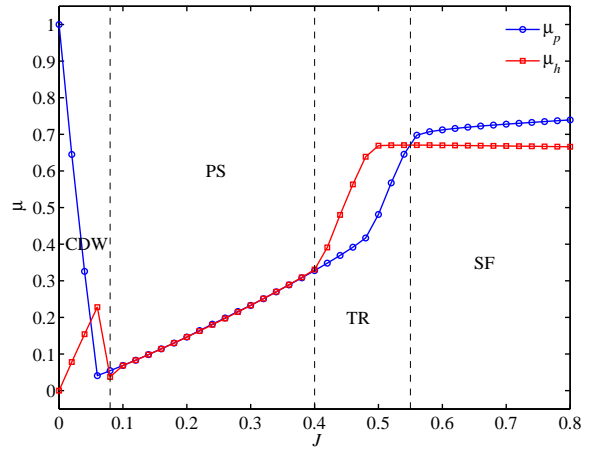}
\caption{(Color online) Ground state phase diagram of the Hamiltonian (\ref{eq1}) with $W=-U$, $N=30$ and $L=59$. Between the CDW and SF phases there exists an exotic PS phase. The crossover (denoted as TR) between PS and SF phases is due to the finite size effect.} \label{fig3}
\end{center}
\end{figure}

\begin{figure}[tbp]
\begin{center}
\includegraphics[width=0.9\columnwidth]{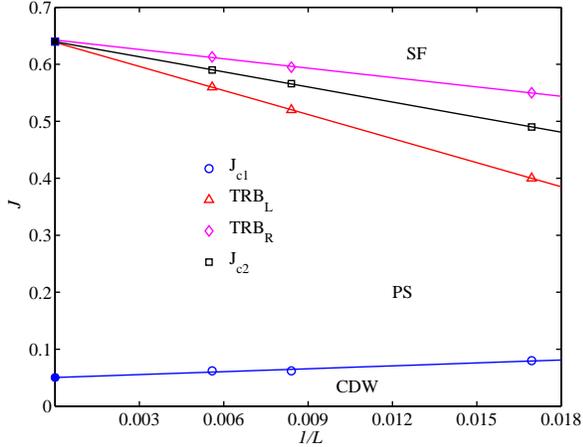}
\caption{(Color online) Finite size scaling for the phase transition points $J_{c1}$ and $J_{c2}$. The result is obtained by three system sizes $L=59,119,179$. Red triangles and magenta diamonds denotes the left and right boundary of the transition region from PS to SF phases, respectively. The solid square and circle are extrapolated results of $J_{c1}$ and $J_{c2}$, respectively. } \label{fig4}
\end{center}
\end{figure}

In the following we present the numerical results for the interesting case of $W = -U$ and take $U$ as the unit of energy. Fig. \ref{fig3} shows the chemical potentials as a function of the hopping rate $J$ for $N = 30$ and $L = 59$. In contrast to the pure repulsive cases, the chemical potentials show anomalous behaviors. Apparently, the phase diagram can be divided into three regions, one is the narrow solid phase, which is the known phase in the literature, the second is the superfluid phase at large hopping rate shown in the right-hand side of the plot. The most strange is the intermediate region where the particle and hole chemical potentials merge. This is an exotic phase separation phase, as discussed in detail below. Therefore, the present system consists of three phases, between the conventional solid and superfluid phases, there exists an exotic phase separation phase and thus there should exist two transition points. Due to finite size effect, the transition from PS to SF phases shows as a crossover, as denoted as the transition region (TR) between the two perpendicular dashed lines on the right-hand side of Fig. \ref{fig3}. In Fig. \ref{fig4} we linearly fit the phase transition points by finite size scaling and find that $J_{c1} \sim 0.05$ and $J_{c2} \sim 0.64$. The transition region disappears in the thermal dynamic limit ($1/L\rightarrow\infty$). The determination method of $J_{c1}$ and $J_{c2}$ are demonstrated in the following two subsections.

To identify these phases, in the following we consider in detail the characteristic features of the real space density distribution, the structure factor of density-density correlation functions, the three-body correlation functions and information entropy.

\subsection{Density distribution and correlation functions}

\begin{figure}[tbp]
\begin{center}
\includegraphics[width=\columnwidth]{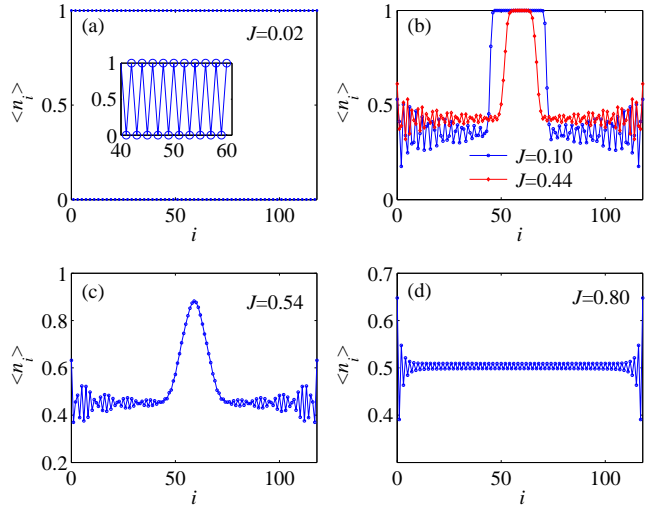}
\caption{(Color online) Boson distribution in real space at different phases or regions. (a) Solid phase, (b) Phase separation phase, (c) Transition region, and (d) Superfluid phase. The inset in (a) is expanded distribution. The parameters used are $N = 60$ and $L = 119$.}\label{fig5}
\end{center}
\end{figure}

Fig. \ref{fig5} shows the real space density distribution of bosons for different hopping rate. As $J$ is quite small, the system is in solid phase and the bosons occupy alternatively the lattice sites, as shown in Fig. \ref{fig5}(a) and the inset in it. Increasing $J$, the bosons trend to hop among the lattice sites. However, due to the interplay between the two-body repulsive and three-body attractive interactions, once the solid phase becomes unstable, the bosons like to form an fully occupied island in order to lower the potential energy. As pointed out above, the island has a potential energy of $U$ neglecting its length. However, obviously this island with energy $U$ is unstable and the bosons at the two ends of it begin to hop outward in order to lower the energy. Meanwhile, the bosons inside the island still keep unmoved due to the hard-core constraint. This result leads to the exotic phase separation phase, as shown in Fig. \ref{fig5}(b). It consists of localized bosons in the middle and delocalized bosons in the rest of real space behaves as liquid. Further increasing $J$, the bosons island gradually shrinks, even to a broad peak [see Fig. \ref{fig5}(c)]. Finally, as $J$ becomes large and dominates the interactions, the system enters into the superfluid phase, the occupied island eventually disappears and the bosons almost distribute uniformly [see Fig. \ref{fig5}(d)].

To further clarify these phases, we calculate the density-density correlation function $\langle n_i n_j\rangle$ and consider its structure factor
\begin{equation}
S_{CDW}(k)=\frac{1}{L^2}\sum_{i,j}\exp[ik(i-j)]\langle n_i n_j\rangle.
\end{equation}

\begin{figure}[tbp]
\begin{center}
\includegraphics[width=\columnwidth]{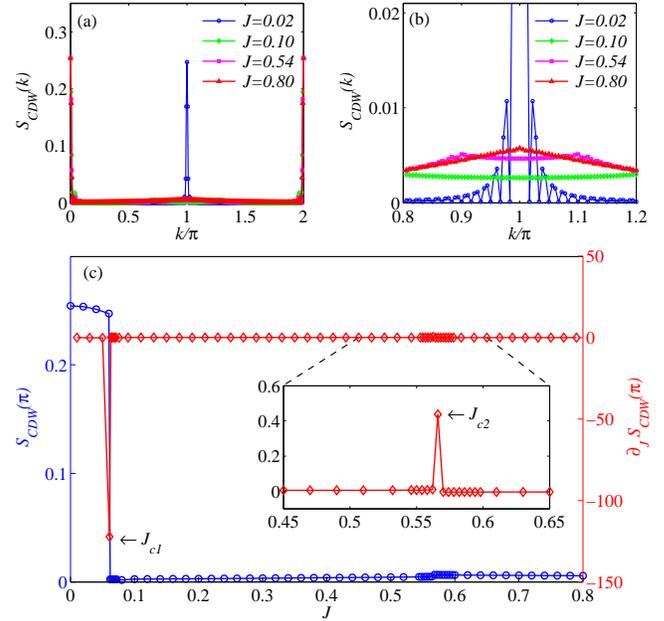}
\caption{(Color online) The behavior of the structure factor $S_{CDW}(k)$. (a) $S_{CDW}(k)$ as a function of $k$ in momentum space and (b) the expanded one around $k = \pi$. (c) $S_{CDW}(\pi)$(blue circle-line, left axis) and its derivation (red diamond-line, right axis) with respect to $J$. Two sharp peaks of $\partial_J S_{CDW}(\pi)$ indicate the positions of the transition points. The parameters used are the same as those in Fig. \ref{fig5}. } \label{fig6}
\end{center}
\end{figure}

Fig. \ref{fig6}(a) shows the structure factor for different phases. Fig. \ref{fig6}(b) is the expanded one around $k = \pi$. In solid phase with $J=0.02$, there is a sharp peak at $k=\pi$. In phase separation phase ($J=0.10,0.54$), $S_{CDW}(k)$ around $\pi$ shows featureless structure, which is different to the broad peak in superfluid phase with $J=0.80$. Fig. \ref{fig6}(c) summarizes the structure factor at $k=\pi$ as a function of $J$. There are two transitions to happen, one is around $J_{c1}=0.06$, where the structure factor peak suddenly drops, indicating a transition from solid phase to the phase separation phase. The other is around $J_{c2}=0.57$, where the phase separation phase goes across the superfluid phase. These transitions can be clearly identified by the derivation of the structure factor as a function of $J$, as shown as red diamond-line where two peaks occur.

\begin{figure}[tbp]
\begin{center}
\includegraphics[width=0.9\columnwidth]{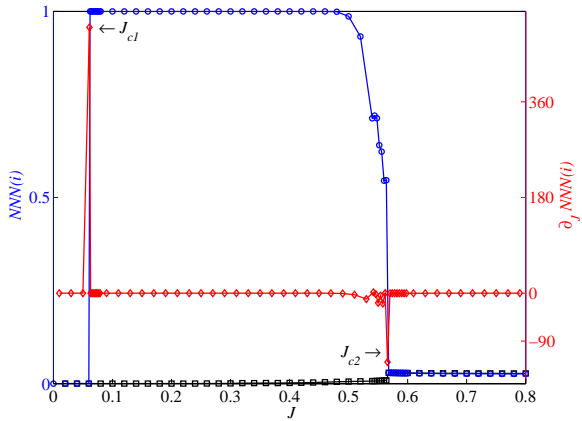}
\caption{(Color online) The three-body correlation function as a function of $J$. The blue circle-line is for $i = (L+1)/2$, the position at the center of the occupied sites and the black square-line is for $i = (L+1)/4$, the position far from the center and the boundary. The red diamond-line denotes the derivation of the $NNN((L+1)/2)$ with respect to $J$. Two phase transition points are indicated by two peaks. The parameters used are the same as those in Fig. \ref{fig5}.} \label{fig7}
\end{center}
\end{figure}

It is also interesting to consider the three-body correlation function $NNN(i)= \langle n_{i-1} n_{i} n_{i+1} \rangle$. In Fig. \ref{fig7} the blue circle-line represents the correlation function of $i = (L+1)/2$ with respect to $J$. In small $J$, no three-boson cluster exists and the correlation function is strictly zero. However, in the phase separation region, the fully occupied boson island forms and the three-body correlation function in the middle of the chain is $1$. Increasing $J$ up to $J_{c2}$, NNN((L+1)/2) dropping to a finite value indicates the entrance of superfluid phase. The black square-line in Fig. \ref{fig7} is the three-body correlation function at $i=(L+1)/4$, where far away from the fully occupied boson island and the edges of the lattice. Its low value between $J_{c1}$ and $J_{c2}$ indicates that bosons outside the boson island behaves as dilute liquid in PS region.

In a brief summary, through the analysis of the characteristic behaviors of the real space density distribution, the structure factor of the density-density correlation function and the three-body correlation function, we confirm that the interplay of the two-body repulsive and three-body attractive interactions and the hopping rate can lead to exotic PS phase between the CDW and SF phase and two phase transition points can be identified by all above physical observables. In the following we further discuss these phase transitions from the point of view of quantum information entropy.

\subsection{Quantum information entropy}
\begin{figure}[tbp]
\begin{center}
\includegraphics[width=0.9\columnwidth]{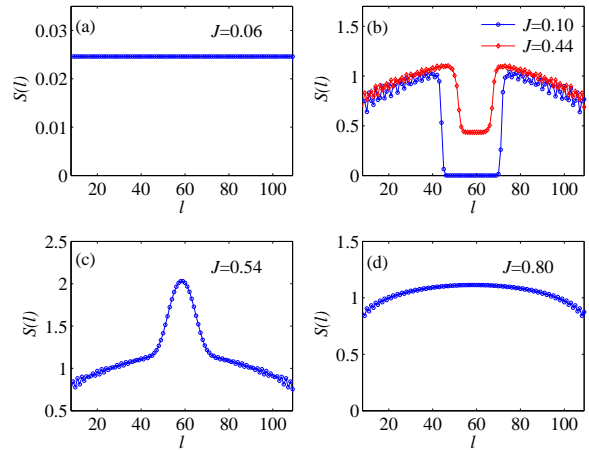}
\caption{(Color online) Block entropy $S(l)$ as a function of the lattice site $l$ for different phases: (a) solid phase, (b) phase separation phase, (c) transition region, (d) superfluid phase. The parameters used are the same as those in Fig. \ref{fig5}.}\label{fig8}
\end{center}
\end{figure}

In recent years, the phase transition physics has been intensively explored by quantum information entropy concept, by which the transition between the Mott insulating and the superfluid phase in Bose-Hubbard model has been clearly identified with repulsive two-body \cite{Buonsante2007} or three-body interaction. \cite{Valencia2012} Here it is also interesting to use the entropy concept to analyze the exotic phases in the present model. Here we use the von Neumann entropy, which is defined by $S_L(A) = \text{Tr} \rho_A \ln \rho_A$, where $\rho_A$ is the reduced density matrix, $\rho_A = \text{Tr}_B \rho$ and $\rho$ is the density matrix of the whole system including two parts $A$ and $B$. For open boundary condition,\cite{Calbrese2004, Valencia2012} the von Neumann entropy saturates (diverges) if the system is gapped (gapless), namely
\begin{equation}\label{eq10}
S_L(l) =\left\{
\begin{array}{rcl}
(c/6) \ln[(L/\pi) \sin(\pi l /L) + \theta, \text{gapless}£¬\,\, (a)\\
(c/6) \ln[\xi_L] + \theta', \text{gapped}\,\, (b)
\end{array} \right.
\end{equation}
where $\theta$ and $\theta '$ are non universal constants, $c$ is the central charge and $\xi_L$ is the correlation length. The behavior of the block entropy $S(l)$ is shown in Fig. \ref{fig8}. For $J = 0.06$ [see Fig. \ref{fig8}(a)], the block entropy is independent of block size $l$, showing that the system is in a solid phase with a finite correlation length, which is the case of Eq. (\ref{eq10})(b). When $J$ increases, the system enters into the phase separation region [see Fig. \ref{fig8}(b)], two liquid parts are divided by a fully occupied boson island in the middle of the chain. If dividing the system at this fully occupied solid bond, the left and right blocks seem to be two isolated parts, with no entanglement and $S(l) = 0$ (see the case of $J=0.1$). For a larger $J=0.44$, the system is still in phase separation region, but the block entropy is nonzero, because the size of boson island becomes smaller than correlation length in liquid phase. There exits a transition region between phase separation and superfluid phases as shown in Fig. \ref{fig8}(c), where two separated fluid phase has a maxima entanglement entropy. In superfluid phase ($J$=0.8), $S(l)$ behaves as that of gapless liquid, which is consistent with Eq. (\ref{eq10})(a).

To further analyze the block entropy $S_L(l)$, similar to the estimator proposed by Lauchli and Kollath, \cite{Lauchli2008} we define
\begin{equation}
\Delta_S(L) = S_L(L/2) - S_L(L/4).
\end{equation}

\begin{figure}[tbp]
\begin{center}
\includegraphics[width=0.9\columnwidth]{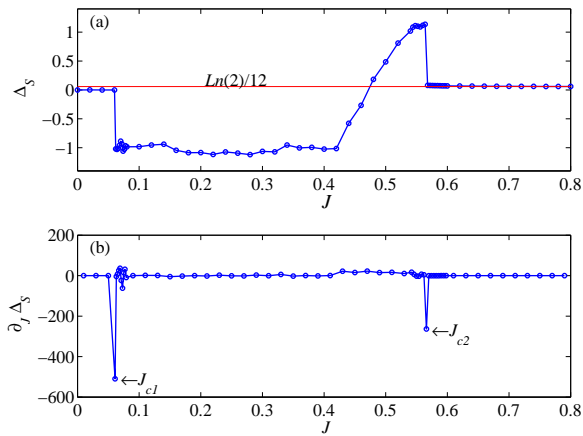}
\caption{(Color online) (a) $\Delta_S$ and (b) its derivation $\partial _J \Delta_S$ as a function of $J$. The critical points $J_{c1}$ and $J_{c2}$ are identified as the positions of sharp jumps of $\partial_J \Delta_S$. The parameters used are the same as those in Fig. \ref{fig5}.} \label{fig9}
\end{center}
\end{figure}

According to Eq. (\ref{eq10}), $\Delta_S(L) = (c/12)\ln(2)$ at gapless phase and $\Delta_S(L)=0$ in solid (gapped) phase.
It is interesting to employ $\Delta_S(L)$ to identify different phases and the phase transition points. In Fig. \ref{fig9}, we plot $\Delta_S$ as a function of $J$. In solid (gapped) phase, $\Delta_S(L) = 0$  and in superfluid phase, $\Delta_S(L) = (c/12)\ln(2)$ (central charge $c=1$ for $J \rightarrow \infty$). In phase separation phase, the block entropy dose not behave as Eq. (\ref{eq10}), $\Delta_S < 0$ is not reasonable in normal phases. In the transition region, $\Delta_S$ is far beyond its value at free boson limit. As $J$ increases to $J_{c2}$, $\Delta_S$ suddenly drops. The phase transition points are determined by the positions of sharp peaks of $\partial _J \Delta_S$ as shown in Fig. \ref{fig8}(b) and are precisely agreed with the results obtained from $\partial_J S_{CDW}(\pi)$ and $\partial_J NNN(i)$ in the preceding subsection.

\section{Summary and discussion}

We have proposed an exotic phase separation phase obtained by the interplay between the repulsive two-body and the attractive three-body interactions and the hopping rate for hard-core bosons in one-dimensional optical lattice. When the interaction strengths are comparable, intuitively physical analysis shows that bosons trend to occupy the sites to form an occupied island where the bosons at the two ends of the occupied island hop outward to lower the energy once the hopping rate is switched on and the solid phase with CDW order becomes unstable. This is the initial state of the exotic phase separation phase. When the hopping rate becomes dominant, the system is always in superfluid phase. This picture has been further confirmed by analyzing the real space density distribution, the structure factor of the density-density correlation functions, the three-body correlation functions and finally quantum information entropy obtained by numerically accurate DMRG calculations. The phase transition points have been precisely determined by the sharp jumps of the first derivative functions of $S_{CDW}(\pi)$, $NNN(i)$ and $\Delta_S$ independently with great accordance. It is expected that these results can be observed in the future cold atom and/or ultra-cold molecule systems due to the ability of independently tuning the interaction parameters in such systems. The experimental realization of these phases can further help to understand the physics of the transition from insulating to superconducting phases, a ubiquitous phenomenon in condensed matter physics.

\section*{Acknowledgments} The work is partly supported by the programs for NSFC, PCSIRT (Grant No. IRT1251), the national program for basic research and the Fundamental Research Funds for the Central Universities of China.


\begin{thebibliography}{99}
\bibitem{note} The choice of OBCs brings in a degenerate ground state in solid phase with CDW order. However, this degeneracy can be eliminated by reducing the lattice size by one. Therefore in the present paper, for half-filling case, we take odd number lattice size $L$ and fix the number of bosons as $N = (L+1)/2$. Likewise, for $2/3$-filling case, we fix the number of bosons as $N = 2(L+1)/3$. In both cases, the correction for the filling factor is about $O(\frac1L)$, which approaches zero in thermodynamic limit.
\end{thebibliography}
\end{document}